\documentclass[aps,prl,preprint,onecolumn,superscriptaddress]{revtex4}
\usepackage[]{graphicx}

\begin{document}

\title{Microswimmers in Patterned Environments}

\author{Giovanni Volpe}
\email{g.volpe@physik.uni-stuttgart.de}
\affiliation{Max-Planck-Institut f\"{u}r Intelligente Systeme, Heisenbergstra{\ss}e 3, 70569 Stuttgart, Germany}
\affiliation{2. Physikalisches Institut, Universit\"{a}t Stuttgart, Pfaffenwaldring 57, 70569 Stuttgart, Germany}

\author{Ivo Buttinoni}
\affiliation{2. Physikalisches Institut, Universit\"{a}t Stuttgart, Pfaffenwaldring 57, 70569 Stuttgart, Germany}

\author{Dominik Vogt}
\affiliation{2. Physikalisches Institut, Universit\"{a}t Stuttgart, Pfaffenwaldring 57, 70569 Stuttgart, Germany}

\author{Hans-J\"{u}rgen K\"{u}mmerer}
\affiliation{2. Physikalisches Institut, Universit\"{a}t Stuttgart, Pfaffenwaldring 57, 70569 Stuttgart, Germany}

\author{Clemens Bechinger}
\email{c.bechinger@physik.uni-stuttgart.de}
\affiliation{Max-Planck-Institut f\"{u}r Intelligente Systeme, Heisenbergstra{\ss}e 3, 70569 Stuttgart, Germany}
\affiliation{2. Physikalisches Institut, Universit\"{a}t Stuttgart, Pfaffenwaldring 57, 70569 Stuttgart, Germany}

\date{\today}

\begin{abstract}
We demonstrate with experiments and simulations how microscopic self-propelled particles navigate through environments presenting complex spatial features, which mimic the conditions inside cells, living organisms and future lab-on-a-chip devices. In particular, we show that, in the presence of periodic obstacles, microswimmers can steer even perpendicularly to an applied force. Since such behaviour is very sensitive to the details of their specific swimming style, it can be employed to develop advanced sorting, classification and dialysis techniques.
\end{abstract}


\maketitle

Brownian motion is the result of random collisions between a microscopic particle and the molecules of the surrounding fluid. In contrast, self-propelled particles additionally take up energy from their environment and convert it into directed motion \cite{r1,r2,r3}; accordingly, their motion is a superposition of random fluctuations and active swimming. Examples of such microswimmers range from chemotactic cells \cite{r4,r5,r6} to artificial systems, where, e.g., artificial flagella are put into motion by magnetic fields \cite{r7,r8,r9,r10} or thermophoretic forces \cite{r11}; also, micron-sized Janus particles have been realized where partial coating with a catalyst leads to non-isotropic electrochemical reactions and thus to directed motion \cite{r12,r13,r14}. Until now most studies have concentrated on the behaviour of microswimmers in homogeneous environments, where one typically observes a crossover from ballistic motion at short times to enhanced diffusion at long times, the latter due to random changes in the swimming direction \cite{r15,r16,r17}. However, self-propelled particles often move in patterned environments, e.g., inside the intestinal tract, which provides the natural habitat of \emph{E. coli} \cite{r5}, or during bioremediation, where chemotactic bacteria spread through porous polluted soils \cite{r18}. In a similar fashion, artificial microswimmers must also reliably perform their tasks in complex surroundings, e.g., inside lab-on-a chip devices \cite{r19} or living organisms. \\
\begin{figure}
\includegraphics[width=8.7cm]{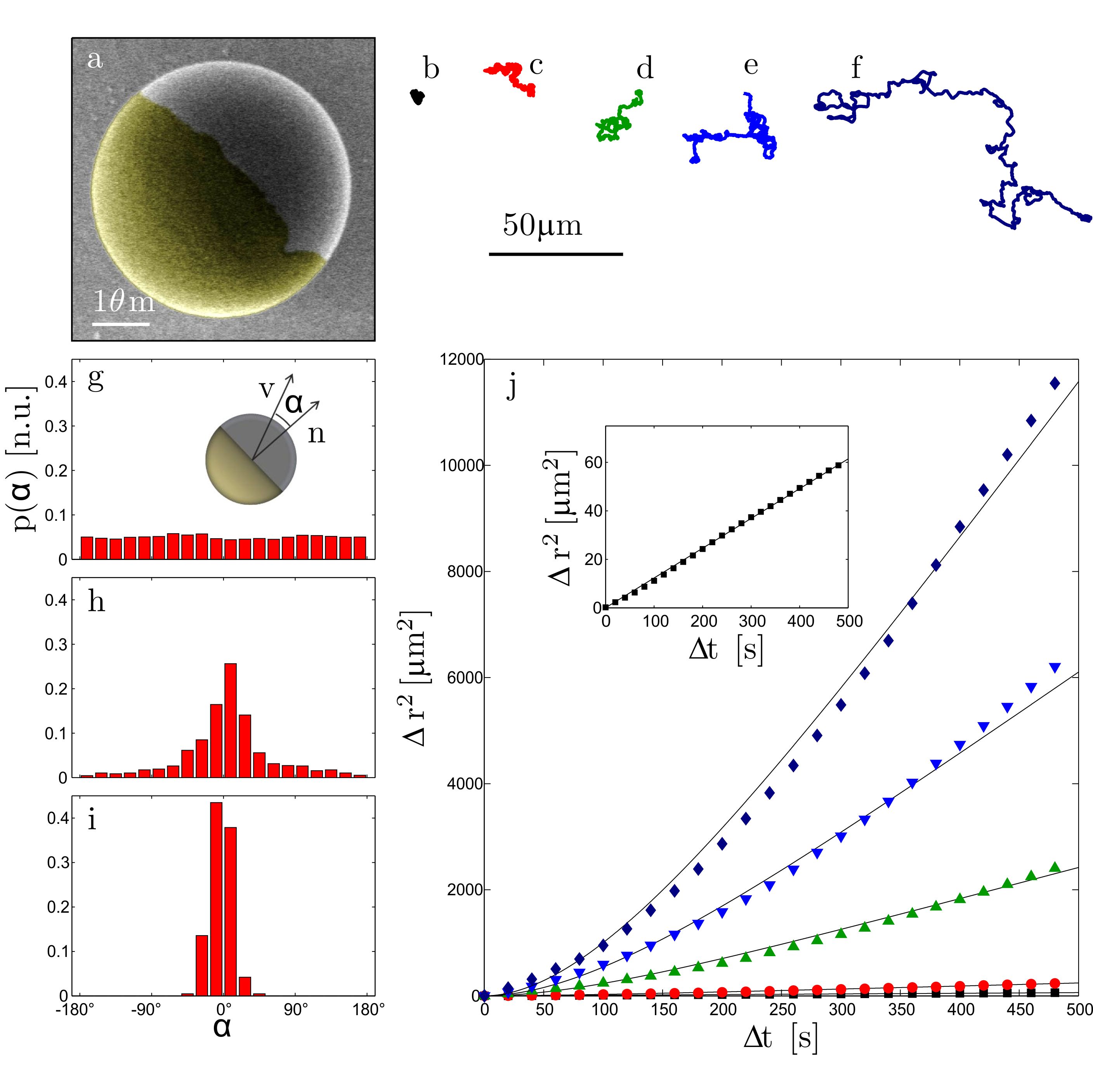}
\caption{(color online) Self-propulsion of Janus particles in critical mixtures. (a) Scanning electron microscopy image of a colloidal particle with a $\mathrm{20\, nm}$ thick gold cap (highlighted). (b-f) 2D trajectories (1000s) of a Janus particle for illumination intensities $I = 0,\, 69,\, 92,\, 115,\, 161\, \mathrm{nW/\mu m^2}$ (respectively to b, c, d, e and f). (g-i) Probability distribution of the angle between the cap and the particle velocity for $I = 0$ (g), $69$ (i), and $138\, \mathrm{nW/\mu m^2}$ (i). (j) Experimentally determined mean square displacements (symbols) and relative fittings to Eq. (1) (lines) for $I = 0,\, 69,\, 92,\, 115,\, 161\, \mathrm{nW/\mu m^2}$ (respectively squares, circles, up-pointing triangles, down-pointing triangles and diamonds). The inset is a magnification of the data for $I=0$. The solid lines correspond to fits according to Eq. (1), see also Tab. 1.\label{fig1}}
\end{figure}
\begin{figure*}
\includegraphics[width=17.4cm]{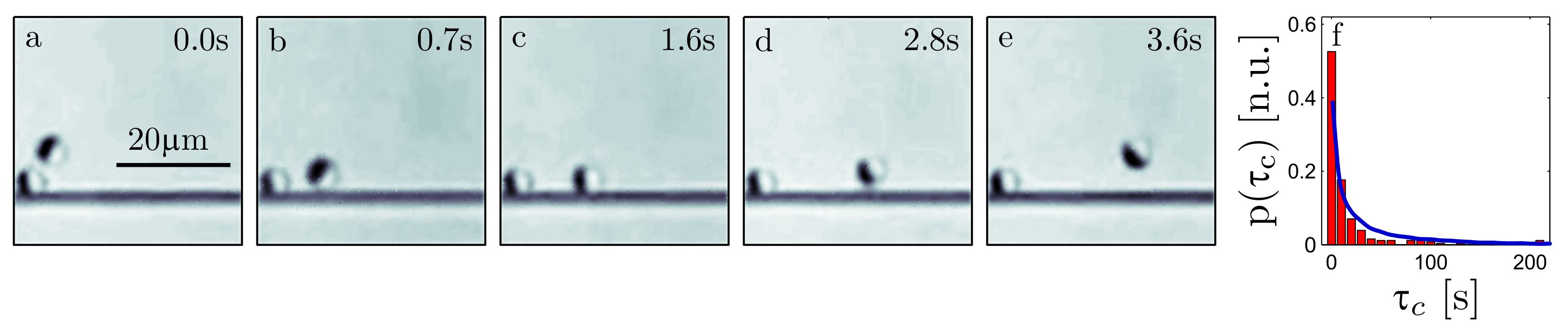}
\caption{(color online) Encounter between a microswimmer and a wall. (a-e) Time series of snapshots demonstrating the approach (a), contact (b-d) and detachment (e) of a Janus particle and a wall. The particle in the left corner is an irreversibly stuck particle which serves as a reference position. (f) Experimentally measured (bars) particle-wall contact time distribution $p(\tau_c)$ between a microswimmer and a wall ($I=115\, \mathrm{nW/\mu m^2}$) for more than 250 encounter events. The line represents the result of a numerical simulation with $L$ and $\tau$ taken from the experimental values of Tab. 1.\label{fig2}}
\end{figure*}
Here, as a first step towards more realistic conditions under which such microswimmers are found and will be employed, we study their behaviour in patterned surroundings where frequent encounters with obstacles become important.  In particular, we investigate their motion within environments featuring simple topographical structures such as a straight wall and periodically arranged obstacles. Self-propelled Janus particles (Fig. 1a) are obtained from paramagnetic silica spheres with radius $R=2.13\, \mathrm{\mu m}$ (Microparticles, SiO2-MAG-S1975). A layer of such particles is  deposited onto a substrate. After the solvent evaporated, their upper sides are coated by thermal evaporation with $2\, \mathrm{nm}$ chromium and $20\, \mathrm{nm}$ gold. This leads to the formation of capped particles. The gold surface is functionalized with COOH-terminated thiols (11-Mercapto-undecanoic acid) to render the caps strongly hydrophilic. When such Janus particles are suspended in a critical mixture of water and 2,6-lutidine\cite{r20} (0.286 mass per cent of lutidine) below the critical temperature $T_C = 307\, \mathrm{K}$, they undergo normal Brownian motion (Fig.1b). However, when the entire sample cell is homogeneously illuminated with light ($\lambda = 532\, \mathrm{nm}$) at low intensities $I$ ($< 0.2\, \mathrm{\mu W/\mu m^2}$), the particle's motion strongly depends on the incident light intensity as shown in Figs. 1c-f. In particular the trajectories become ballistic at short times as this is typically found for self-propelled objects. To follow the particles' trajectories with video microscopy, their motion is vertically confined by adjusting the sample height to about $7\, \mathrm{\mu m}$. Such sample cells are made from two microscope cover slides. The porous structures and the walls, which also work as spacers, are realized on top of the first slide with soft lithography where a $7\, \mathrm{\mu m}$ thick layer of photoresist (MicroChem, SU-8 2005) is first spin-coated and then exposed to ultraviolet light through appropriate transmission masks. After development (Micro Resist Techn. GmbH, mr Dev 600) and hard backing the structures are filled with colloidal suspensions and sealed with the second cover slide and epoxy glue.
\begin{table}
\caption{\label{Tab1} Characterization of self-propulsion. From the fit of Eq. (1) to the measured MSD we obtain the intensity dependent average swimming speed $v$ and time $\tau$ during which the swimming direction is maintained. From $v$ and $\tau$ we calculate the swimming length $L=v\tau$ and the effective diffusion coefficient $D_{eff} = D_0 + \frac{L^2}{4\tau}$. The errors correspond to standard deviations.}
\begin{ruledtabular}
\begin{tabular}{ccccc}
$I$ ($ \mathrm{nW/\mu m^2}$) & $\tau$ ($ \mathrm{s}$) & $v$ ($ \mathrm{nm/s}$) & $L$ ($ \mathrm{\mu m}$) & $D_{\mathrm{eff}}$ ($ \mathrm{\mu m^2/s}$)\\
$0$ & $-$ & $-$ & $-$ & $0.031 \pm 0.006$\\
$46$ & $220 \pm 20$ & $46 \pm 5$ & $10 \pm 2$ & $0.147 \pm 0.029$\\
$69$ & $190 \pm 20$ & $85 \pm 7$ & $16 \pm 3$ & $0.370 \pm 0.069$\\
$92$ & $190 \pm 20$ & $175 \pm 20$ & $33 \pm 8$ & $1.49 \pm 0.42$\\
$115$ & $220 \pm 30$ & $265 \pm 33$ & $58 \pm 14$ & $3.89 \pm 1.01$\\
$138$ & $240 \pm 40$ & $310 \pm 28$ & $74 \pm 13$ & $5.80 \pm 1.49$\\
$161$ & $230 \pm 40$ & $360 \pm 27$ & $83 \pm 12$ & $7.48 \pm 1.86$\\
\end{tabular}
\end{ruledtabular}
\end{table}
The propulsion is caused by diffusiophoresis, which occurs when a particle is subjected to a concentration gradient within the solvent \cite{r21}. Here, such gradients are created by the particle itself (self-diffusiophoresis \cite{r22,r23}) because the incident light is absorbed by the gold cap and thus leads to a local demixing of the binary solvent once the cap temperature exceeds $T_C$. In case of a hydrophilic cap this leads to an accumulation of the water-rich phase at the gold cap.\\
From the recorded videos, we also determined the particleÕs time-dependent propulsion direction $\mathbf{v}$ together with their current cap orientation $\mathbf{n}$ and the enclosed angle $\alpha$ (see inset of Fig. 1g). Figs. 1g-j show the normalized probability distribution $p(\alpha)$ for increasing illumination intensity $I$. Clearly, $p(\alpha)$ becomes strongly peaked around $\alpha = 0$ with growing $I$ which demonstrates that the propulsion force acts in the direction opposite to the gold cap. When the gold cap is made hydrophobic, the propulsion direction becomes reversed since then the lutidine-rich phase accumulates at the cap and thus leads to a change in the sign of the concentration gradient.\\
The symbols in Fig.1j show the experimentally determined mean square displacements (MSD) for the conditions of Figs.1b-f. In the absence of illumination the particles undergo normal Brownian motion, where the MSD equals $\Delta r^2 = 4D_0\Delta t$ (inset of Fig.1j) with diffusion coefficient $D_0=0.031\pm 0.006\,  \mathrm{\mu m^2/s}$. With increasing intensity, the MSDs clearly deviate from a diffusive behaviour but are well described by 
\begin{equation}
\Delta r^2 = \left[ 4 D_0 + \frac{L^2}{\tau} \right] \Delta t + \frac{L^2}{2} \left[ \exp\left(-\frac{2\Delta t}{\tau}\right) - 1 \right],
\end{equation}
which generally characterizes the motion of self-propelled particles \cite{r15,r17}. Here, $\tau$ is the average timescale over which the trajectory direction is maintained. Accordingly, the average velocity of the particle is $v = L/\tau$ \cite{r15} with $L$ the swimming length, i.e. the average length of rather straight segments in the particleÕs trajectory. Due to the random changes in the particleÕs direction, for long times ($\Delta t >> \tau$) Eq. (1) leads to an effective diffusion with $D_{eff} = D_0 + \frac{L^2}{4\tau}$, while for short times ($\Delta t << \tau$) the particleÕs motion becomes ballistic with $\Delta r^2 \propto L^2 \Delta t^2$. The values of the free fitting parameters $\tau$ and $v$ for which best agreement with the data has been achieved are listed together with $L$ and $D_{\mathrm{eff}}$ Tab.1. From this it becomes obvious that $L$ increases (above some threshold value) linear with the illumination intensity. In contrast, no significant variation of $\tau$ with $I$ is observed. The value $\tau \approx 200\,  \mathrm{s}$ for which best agreement with the experimental data is obtained is close to the timescale of rotational diffusion $\tau_R=1/D_R=4R^2/3D_0$ with $D_R$ the rotational diffusion coefficient. This yields $\tau_R=188\,  \mathrm{s}$ for the $R=2.13\,  \mathrm{\mu m}$ particles which is indeed close to $\tau$. Similar agreement (within $\pm10\%$) between $\tau$ and $\tau_R$ has also been found for Janus particle with radii between $R=1.0$ and $0.5\,  \mathrm{\mu m}$. This suggests that directional changes in the trajectories result from cap reorientations due to rotational diffusion.\\
\begin{figure}
\includegraphics[width=8.7cm]{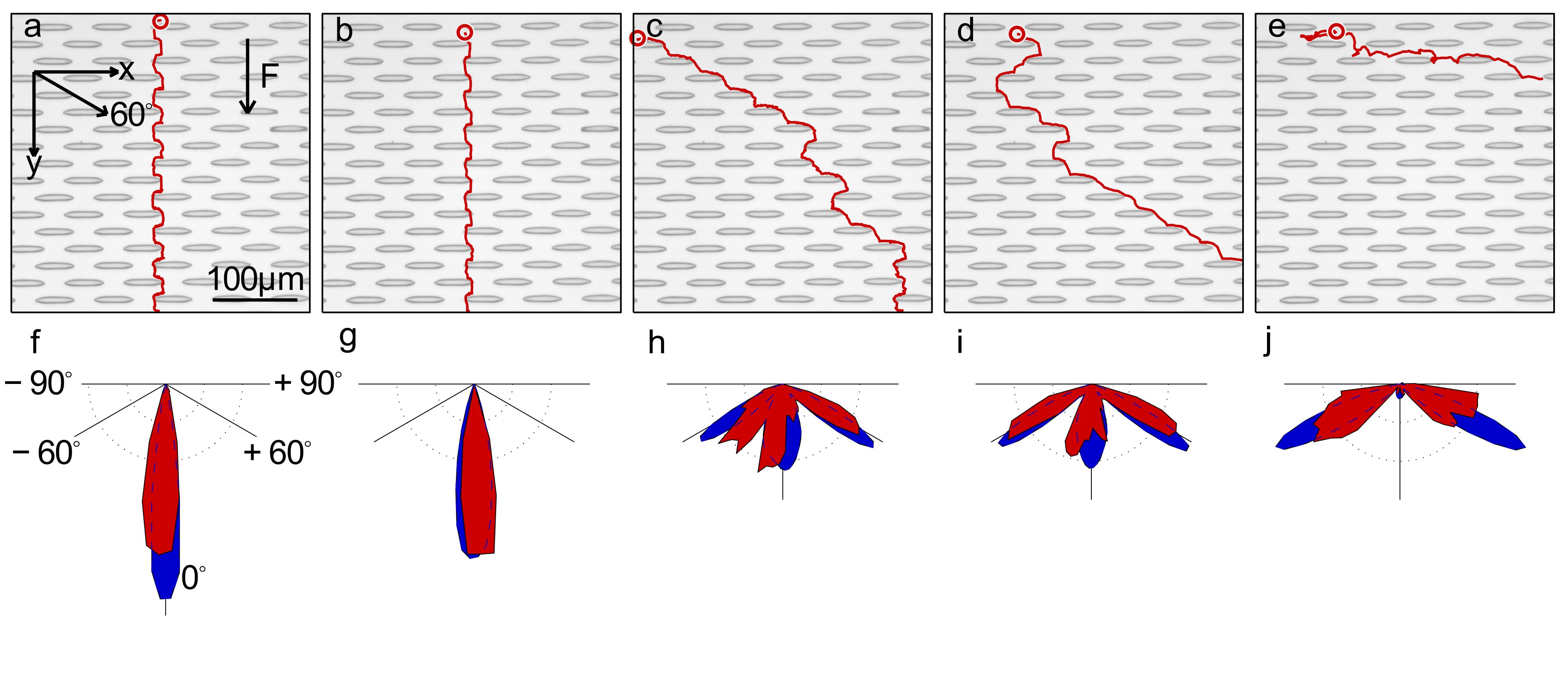}
\caption{(color online) Microswimmers in a patterned environment. (a-e) Typical trajectories of self-propelled particles moving through a triangular lattice (lattice constant $L_c = 35\,\mathrm{\mu m}$) of elliptical obstacles when a drift force $F=0.12\, \mathrm{pN}$ is applied along the y-direction. (a) Brownian particle (no propulsion), (b) $L = 16\, \mathrm{\mu m}$, (c) $L = 24\, \mathrm{\mu m}$, (d) $L = 33\, \mathrm{\mu m}$ and (e) $L = 83\, \mathrm{\mu m}$. (f-j) Corresponding histograms of the experimentally measured (red/light grey) and simulated (blue/dark grey) directions of the particle trajectories as defined by two points in the trajectory separated by $100\, \mathrm{\mu m}$. The parameters for the simulations are taken from Tab. 1.\label{fig3}}
\end{figure}
When self-propelled particles swim through a patterned medium, in particular for high swimming velocities, frequent encounters with obstacles will occur. Figs. 2a-e show the interaction between a Janus particle and a wall. First the particle approaches the wall (Fig. 2a) and gets in contact (Fig. 2b). Then it slides along the wall (Fig. 2c) until rotational diffusion realigns the particle so that its orientation vector $\mathbf{n}$ points away from the wall and leads to particle detachment (Figs. 2d,e). Measuring the distribution of the particle-wall contact time $\tau_c$, we find a monotonic decrease as shown in Fig. 2f (bars). We also performed numerical Brownian dynamics simulations \cite{r29}, where the motion of the Janus particle is modelled by a superposition of random diffusion and ballistic motion with velocity $v$ as inferred from Tab.1. The direction of the ballistic component is determined by the orientation of the particle, which is continuously realigned due to rotational diffusion. Whenever the particle encounters a wall, the translational motion perpendicular to the wall is set to zero, while the horizontal component remains unchanged. From the simulated particle trajectories we calculated the corresponding distribution $p(\tau_c)$ (line in Fig. 2f), which shows good agreement with the experimental data. This suggests that the particle-wall encounter mechanism is correctly described by this simple model and, in particular, that the rotational diffusion remains largely unaffected by the proximity to the wall.\\
We have also investigated the behaviour of self-propelled particles in the presence of a two-dimensional periodic pattern where straight unlimited swims are only possible along certain directions. We have chosen a structure made of a series of ellipsoidal pillars arranged in a triangular lattice (lattice constant $L_c=35\, \mathrm{\mu m}$, Fig. 3a). Within such structures, long swimming cycles are only possible along two main directions: at $\pm 60^{\circ}$ and $\pm 90^{\circ}$ with respect to the y-axis. Otherwise the motion is strongly hindered due to collisions with the obstacles. In the presence of an additionally applied drift force, this leads to strong differences in the particle trajectories depending on their swimming length. In our experiments a constant drift force acting on the paramagnetic particles in the y-direction is generated by a magnetic field gradient. From the average drift speed of a Brownian particle ($I=0\, \mathrm{\mu W/\mu m^2}$), the drift velocity has been determined to $v_d=0.97\, \mathrm{\mu m/s}$, which corresponds to a Stokes force of $F=0.12\, \mathrm{pN}$.\\
The typical trajectory of a Brownian particle is shown in Fig. 3a. Because of the P\'eclet number $Pe \approx 1000$, the effect of the diffusion is rather weak and the particle meanders almost deterministically through the structure in the direction of $F$. For increasing swimming lengths, however, significant changes in the shape of the trajectories are observed. This becomes particularly pronounced for $L>L_c$, where the particles perform swimming cycles of increasing length along the diagonal channels (Figs. 3c,d). For $L=83\, \mathrm{\mu m}$ the propulsion becomes so strong that the particles partially move perpendicular to the drift force (Fig. 3e); occasionally even motion against the drift force can be observed.\\
The direction of the particle motion through the structure is characterized by the direction (with respect to the y-axis) of the line connecting points of the trajectory separated by a distance of $100\, \mathrm{\mu m}$. The probability distributions of these angles are shown by the red/light grey polar histograms in Figs. 3f-j. One clearly observes that with increasing $L$ the propagation of particles along the direction of the applied drift becomes less likely, while trajectories along $\pm 60^{\circ}$, i.e. along the directions that permit long swimming events, become more frequent. We remark that, differently from the deflection of Brownian particles in a periodic potential \cite{r24,r25,r26}, this mechanism relies on the dynamical properties of the microswimmers. We also compare these results with numerical simulations (blue/dark grey polar histograms in Figs. 3f-j), which show good agreement with the experimental data.\\
\begin{figure}
\includegraphics[width=8.7cm]{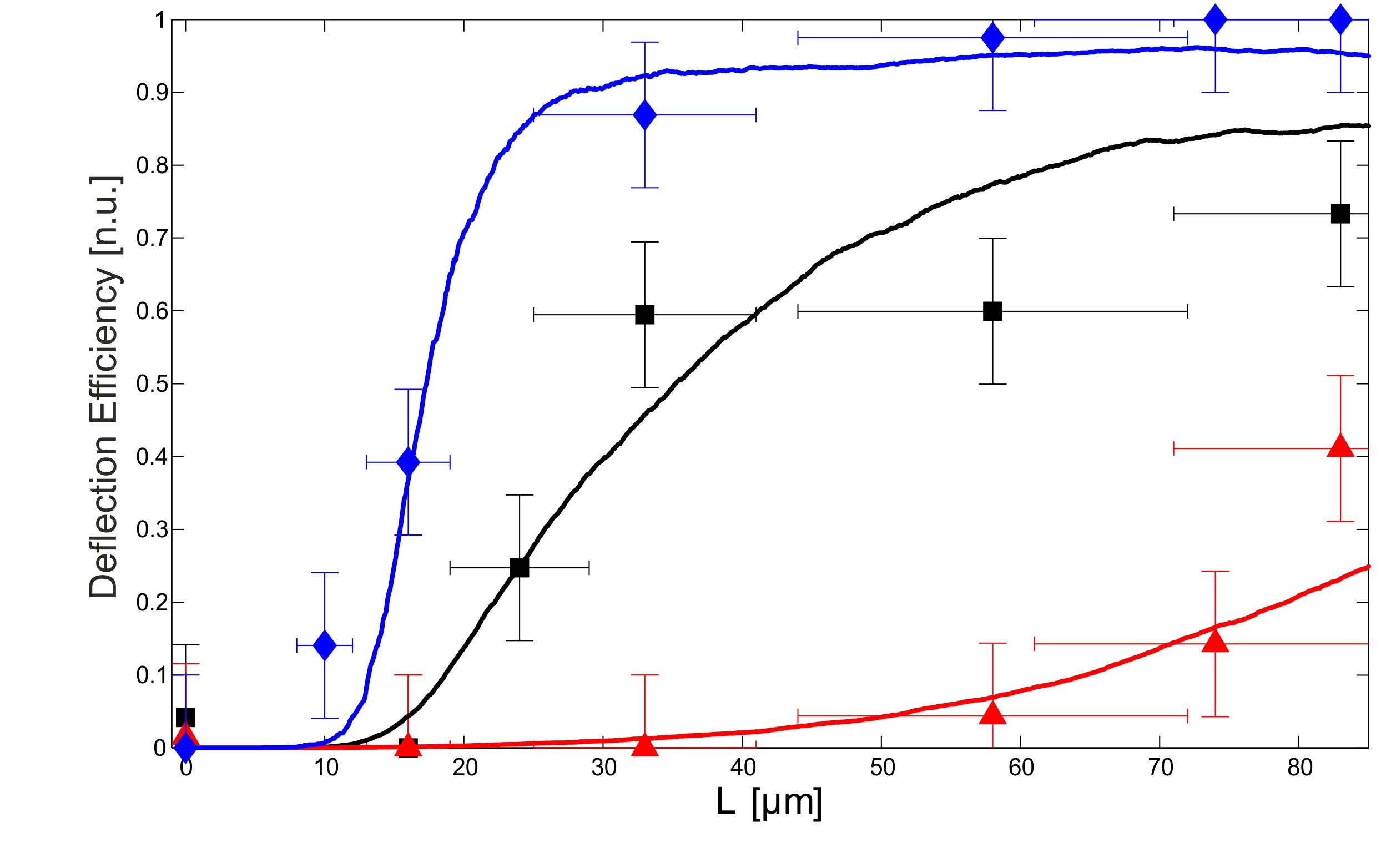}
\caption{(color online) Deflection efficiency. Measured probability that particles are deflected by more than $30^{\circ}$ after a travelling length of $100\, \mathrm{\mu m}$ as a function of the swimming length $L$ for various imposed magnetic drift forces $F=0.06\pm0.02\, \mathrm{pN}$ (diamonds), $0.12\pm 0.05\, \mathrm{pN}$ (squares) and $0.28\pm 0.12\, \mathrm{pN}$ (triangles). The solid lines are the results of numerical calculations. The parameters for the simulations are taken from Tab. 1.\label{fig4}}
\end{figure}
With the additional possibility of varying the drift force, these observations can be exploited to spatially separate self-propelled particles with small differences in their individual swimming behaviour. This is demonstrated in Fig. 4, where we show the deflection efficiency as a function of $L$, as defined by the probability that the mean particle trajectory is deflected by more than $30^{\circ}$ after a travelling length of $100\, \mathrm{\mu m}$. As symbols (lines) we have plotted experimental (simulation) data obtained for different drift forces. The black data correspond to $F=0.12\, pN$ (Fig. 3). Here the deflection efficiency shows a strong increase around $L \approx 30 \mathrm{\mu m}$ and a flattening towards larger swimming lengths. The blue and red data are obtained for $F=0.06$ and $0.28\, \mathrm{pN}$, respectively, and demonstrate that the sorting efficiency strongly responds to variations in $F$. Accordingly, the deflection (transmission) of self-propelled particles while crossing a patterned structure can be easily tuned by the appropriate choice of $F$.\\
Finally, we remark that the sorting mechanism discussed here can be directly applied to other self-propelled objects. In these cases, drift forces can be created, e.g., by electric fields or by a solvent flow through the device. Therefore, we expect that our method may find wider use as an efficient technique to characterize the behaviour of motile cells and bacteria even when the details of their swimming behaviour differs from those of the particles employes in this work. Compared to other concepts for the sorting of chemotactic bacteria, our strategy avoids the need for the creation of chemical gradients which are difficult to stabilize in time and space \cite{r27,r28}.

\begin{acknowledgments}
We gratefully acknowledge V. Blickle and M. Kollmann for inspiring discussions. This work has been partially financially supported by the Marie Curie-Initial Training Network Comploids, funded by the European Union Seventh Framework Program (FP7).
\end{acknowledgments}


\end{document}